\documentstyle[psfig]{mn}

\newif\ifAMStwofonts
\ifoldfss
  \ifCUPmtlplainloaded \else
    \NewTextAlphabet{textbfit} {cmbxti10} {}
    \NewTextAlphabet{textbfss} {cmssbx10} {}
    \NewMathAlphabet{mathbfit} {cmbxti10} {} % for math mode
    \NewMathAlphabet{mathbfss} {cmssbx10} {} %  "   "    "
  \fi
  \ifAMStwofonts
    \ifCUPmtlplainloaded \else
      \NewSymbolFont{upmath} {eurm10}
      \NewSymbolFont{AMSa} {msam10}
      \NewMathSymbol{\upi}     {0}{upmath}{19}
      \NewMathSymbol{\umu}     {0}{upmath}{16}
      \NewMathSymbol{\upartial}{0}{upmath}{40}
      \NewMathSymbol{\leqslant}{3}{AMSa}{36}
      \NewMathSymbol{\geqslant}{3}{AMSa}{3E}

    \fi
  \fi
\fi % End of OFSS

\ifnfssone
  \newmathalphabet{\mathit}
  \addtoversion{normal}{\mathit}{cmr}{m}{it}
  \addtoversion{bold}{\mathit}{cmr}{bx}{it}
  \newmathalphabet{\mathbfit} % math mode version of \textbfit{..}
  \addtoversion{normal}{\mathbfit}{cmr}{bx}{it}
  \addtoversion{bold}{\mathbfit}{cmr}{bx}{it}
  \newmathalphabet{\mathbfss} % math mode version of \textbfss{..}
  \addtoversion{normal}{\mathbfss}{cmss}{bx}{n}
  \addtoversion{bold}{\mathbfss}{cmss}{bx}{n}
  \ifAMStwofonts
    \ifCUPmtlplainloaded \else
      \UseAMStwoboldmath
      \makeatletter
      \new@mathgroup\upmath@group
      \define@mathgroup\mv@normal\upmath@group{eur}{m}{n}
      \define@mathgroup\mv@bold\upmath@group{eur}{b}{n}
      \edef\UPM{\hexnumber\upmath@group}
      \new@mathgroup\amsa@group
      \define@mathgroup\mv@normal\amsa@group{msa}{m}{n}
      \define@mathgroup\mv@bold\amsa@group{msa}{m}{n}
      \edef\AMSa{\hexnumber\amsa@group}
      \makeatother
      \mathchardef\upi="0\UPM19
      \mathchardef\umu="0\UPM16
      \mathchardef\upartial="0\UPM40
      \mathchardef\leqslant="3\AMSa36
      \mathchardef\geqslant="3\AMSa3E
    \fi
  \fi
\fi % End of NFSS release 1

\ifnfsstwo
  \DeclareMathAlphabet{\mathbfit}{OT1}{cmr}{bx}{it}
  \SetMathAlphabet\mathbfit{bold}{OT1}{cmr}{bx}{it}
  \DeclareMathAlphabet{\mathbfss}{OT1}{cmss}{bx}{n}
  \SetMathAlphabet\mathbfss{bold}{OT1}{cmss}{bx}{n}
  \ifAMStwofonts
    \ifCUPmtlplainloaded \else
      \DeclareSymbolFont{UPM}{U}{eur}{m}{n}
      \SetSymbolFont{UPM}{bold}{U}{eur}{b}{n}
      \DeclareSymbolFont{AMSa}{U}{msa}{m}{n}
      \DeclareMathSymbol{\upi}{0}{UPM}{"19}
      \DeclareMathSymbol{\umu}{0}{UPM}{"16}
      \DeclareMathSymbol{\upartial}{0}{UPM}{"40}
      \DeclareMathSymbol{\leqslant}{3}{AMSa}{"36}
      \DeclareMathSymbol{\geqslant}{3}{AMSa}{"3E}
    \fi
  \fi
\fi % End of NFSS release 2

\ifCUPmtlplainloaded \else
  \ifAMStwofonts \else % If no AMS fonts
    \def\upi{\pi}
    \def\umu{\mu}
    \def\upartial{\partial}
  \fi
\fi
\def\gs{\mathrel{\raise1.16pt\hbox{$>$}\kern-7.0pt %  >
\lower3.06pt\hbox{{$\scriptstyle \sim$}}}}         %  ~
\def\ls{\mathrel{\raise1.16pt\hbox{$<$}\kern-7.0pt %  <
\lower3.06pt\hbox{{$\scriptstyle \sim$}}}}         %  ~

\title{Intrinsic Correlation of Galaxy Shapes:
Implications for Weak Lensing Measurements}
\author[Heavens et al.]{Alan Heavens$^{1}$\thanks{afh@roe.ac.uk}, 
Alexandre Refregier$^{2}$\thanks{ar@ast.cam.ac.uk} \&
Catherine Heymans$^{1}$\thanks{cech@roe.ac.uk} \\
$^1$ Institute for Astronomy, Univ. of Edinburgh, Royal Observatory, 
Blackford Hill,Edinburgh, EH9 3HJ,UK \\
$^2$ Institute of Astronomy, Madingley Road, Cambridge CB3 OHA, UK}

\date{Accepted ---. Received ---; in original form ---.}

\pagerange{\pageref{firstpage}--\pageref{lastpage}}
\pubyear{2000}

\begin{document}

\maketitle

\label{firstpage}

\begin{abstract}
Weak gravitational lensing is now established as a powerful method to
measure mass fluctuations in the universe. It relies on the
measurement of small coherent distortions of the images of background
galaxies. Even low-level correlations in the intrinsic shapes of
galaxies could however produce a significant spurious lensing
signal. These correlations are also interesting in their own right,
since their detection would constrain models of galaxy
formation. Using $3\times 10^{4} - 10^5$ halos found in N-body
simulations, we compute the correlation functions of the intrinsic
ellipticity of spiral galaxies assuming that the disk is perpendicular
to the angular momentum of the dark matter halo. We also consider a
simple model for elliptical galaxies, in which the shape of the dark
matter halo is assumed to be the same as that of the light. For deep
lensing surveys with median redshifts $\sim 1$, we find that intrinsic
correlations of $\sim 10^{-4}$ on angular scales $\theta \sim 0.1-10'$
are generally below the expected lensing signal, and contribute only a
small fraction of the excess signals reported on these scales. On
larger scales we find limits to the intrinsic correlation function at
a level $\sim 10^{-5}$, which gives a (model-dependent) range of
separations for which the intrinsic signal is about an order of
magnitude below the ellipticity correlation function expected from
weak lensing.  Intrinsic correlations are thus negligible on these
scales for dedicated weak lensing surveys.  For wider but shallower
surveys such as SuperCOSMOS, APM and SDSS, we cannot exclude the
possibility that intrinsic correlations could dominate the lensing
signal. We discuss how such surveys could be used to calibrate the
importance of this effect, as well as study spin-spin correlations of
spiral galaxies.
\end{abstract}


\section{Introduction}
\label{intro}
Weak gravitational lensing is now established as a powerful method to
directly measure the distribution of mass in the universe (Gunn 1967,
Blandford et al 1991, Villumsen 1996, Bernardeau 1997, Schneider et al
1998; for recent reviews see Mellier 1999; Kaiser 1999; Bartelmann \&
Schneider 1999). This method is based on the measurement of the
coherent distortions that lensing induces on the observed shapes of
background galaxies. It is routinely used to map the mass of clusters
of galaxies (see Fort \& Mellier 1994, Schneider 1996 for reviews) and
has now been applied to a supercluster of galaxies (Kaiser et
al. 1998) and to galaxy groups (Hoekstra et al. 1999). Recently,
several groups have reported the statistical detection of weak lensing
by large-scale structure (Wittman et al. 2000; van Waerbeke et
al. 2000; Bacon, Refregier \& Ellis 2000; Kaiser, Wilson \& Luppino
2000). These detections offer remarkable prospects for precise
measurements of the mass power spectrum and of cosmological parameters
(Kaiser 1992; Jain \& Seljak 1997; Kamionkowski et al.  1997; Kaiser
1998; Hu \& Tegmark 1998; Van Waerbeke et al. 1998).

A potential limitation of this technique is the correlation of the
intrinsic shapes of galaxies which would produce spurious lensing
signals. These intrinsic shape correlations must therefore be
accounted for in weak lensing surveys. In addition, they are
interesting in their own right as their detection would constrain
models of galaxy formation. Such intrinsic correlations could be
produced by several effects: correlations of torques in random
gaussian fields during linear evolution (e.g. Heavens \& Peacock
1988), the coupling of angular momentum of halos during their
non-linear collapse (e.g. Navarro \& Steinmetz 1997), tidal
interaction of nearby galaxies and interaction of the galaxies with
the gravitational potential from surrounding large scale structures
(West, Villumsen \& Dekel 1991, Tormen 1997).

A calculation of intrinsic shape correlation thus requires an
understanding of the origin and properties of the angular momentum of
galaxies, a problem which has puzzled astrophysicists for over 5
decades (see Efstathiou \& Silk 1983 for a review). Hoyle (1949) was
the first to suggest that it arises from the tidal fields of
neighbouring galaxies. Peebles (1969) examined this theory by computing
the growth rate of angular momentum for a spherical collapse using a
second-order expansion.  Doroshkevich (1970) recognised that galaxy
spin emerges through first-order terms if a non-spherical halo was
considered, and White (1984) showed that the resulting growth rate was
linear in time. The statistics of galaxy spins arising from tidal
torques on density peaks have been studied analytically and using
N-body simulations (Heavens \& Peacock 1988; Catalan \& Theuns 1996;
Barnes \& Efstathiou 1987; Sugerman, Summers \& Kamionkowski 1999; Lee
\& Pen 2000 and references therein).

In this paper, we study the correlation of galaxy shapes, quantify its
impact on weak lensing surveys and assess its detectability using wide
shallow surveys. We concentrate on spiral galaxies, and assume that
their disk is perpendicular to the angular momentum vector of their
halo. We compute the correlation of the angular momenta of halo pairs
found in N-body simulations. This allows us to compute the angular
correlation function of the ellipticity of the galaxies projected on
the sky. We compare this intrinsic ellipticity correlation function to
that expected for weak lensing surveys. In addition, we study its
detectability with present and upcoming wide shallower surveys such as
SuprCOSMOS, APM and SDSS. Studies of intrinsic shape correlations
using analytical techniques will be presented in Crittenden et
al. (2000) and Catelan, Kamionkowski \& Blandford (2000); a very
similar independent numerical study by Croft \& Metzler (2000) has
also been completed, while a detection of intrinsic spin correlations
has been reported by Pen, Lee
\& Seljak (2000).

This paper is organised as follows. In \S\ref{ellip}, we define the
ellipticity of galaxies and the associated correlation function.  In
\S\ref{lensing}, we compute the ellipticity correlation expected for
lensing. We compute that arising from intrinsic shape correlations of
galaxies in \S\ref{intrinsic}. In \S\ref{impact}, we discuss the
impact of the intrinsic correlation on weak lensing measurements and
its detectability with wide shallow surveys. In \S\ref{conclusion}, we
summarise our conclusions.

\section{Ellipticity Correlation Function}
\label{ellip}
Following lensing conventions, we characterise the shape of
the images of galaxies on the sky by defining
the ellipticity vector $\epsilon_{i} =\{\epsilon_{1},\epsilon_{2} \}$ as
\begin{equation}
\label{eq:e_def}
\epsilon_{i} = \frac{a^{2}-b^{2}}{a^{2}+b^{2}}
  \{ \cos 2 \alpha, \sin 2 \alpha \}
\end{equation}
where $a$ and $b$ are the major and minor axes of the galaxy image,
and $\alpha$ is its position angle counter-clockwise from the x-axis.
The ellipticity is independent of the surface brightness profile,
provided only that the projected contours of surface brightness are
elliptical.  The ellipticity component $\epsilon_{1}$ ($\epsilon_{2}$)
corresponds to elongation and compressions along (at 45$^{\circ}$
from) the x-axis. Under a rotation of the coordinate system by an
angle $\varphi$ (counter-clockwise from the original x-axis), the
ellipticity $\epsilon_{i}$ transforms into the rotated ellipticity
$\epsilon_{i}^{r}$ given by
\begin{equation}
\label{eq:e_r}
\epsilon_{i}^{r} = R_{ij}(2 \varphi) \epsilon_{j},
\end{equation}
where the rotation matrix is defined as
\begin{equation}
R_{ij}(2 \varphi) = \left( 
\begin{array}{cc}
\cos(2 \varphi) & \sin(2 \varphi) \\
- \sin(2 \varphi) & \cos(2 \varphi) \\
\end{array} \right).
\end{equation}

\begin{figure}
\psfig{figure=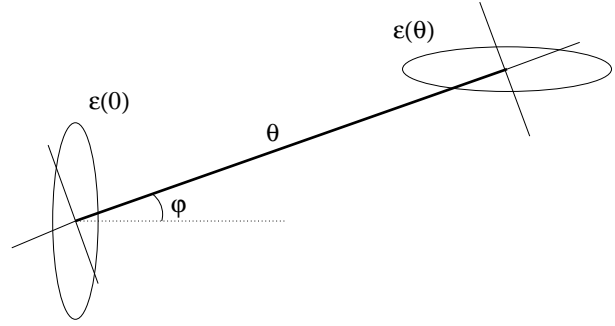,width=80mm} \caption{Geometry of the
ellipticity correlation functions. Two galaxies separated by an angle
$\theta$ are assigned ellipticities $\epsilon_{i}(0)$ and
$\epsilon_{i}(\theta)$. These ellipticities can then be transformed
into the rotated ellipticities $\epsilon_{i}^{r}(0)$ and
$\epsilon_{i}^{r}(\theta)$ defined in the coordinate system (thin
solid lines) which is aligned with the separation vector ${\mathbf
\theta}$. The two correlation functions are then defined as
$C_{1}(\theta) \equiv \langle \epsilon_{1}^{r}(0) \epsilon_{1}^{r}(\theta)
\rangle$ and $C_{2}(\theta) \equiv \langle \epsilon_{2}^{r}(0)
\epsilon_{2}^{r}(\theta) \rangle$}
\label{fig:ecorr}
\end{figure}

The correlation of the shapes of galaxy images can be quantified using
the ellipticity correlation functions. Let us consider two galaxy
images separated by an angular vector ${\mathbf{\theta}}$, with
ellipticities $\epsilon_{i}(0)$ and
$\epsilon_{i}({\mathbf{\theta}})$. The geometry of the correlation
functions is illustrated in Figure~\ref{fig:ecorr}. It is convenient
to consider a coordinate system which is rotated so that its x-axis is
aligned with the separation vector ${\mathbf{\theta}}$. In this
rotated coordinate system, the ellipticities of the two galaxies are
$\epsilon_{i}^{r}(0)$ and $\epsilon_{i}^{r}({\mathbf{\theta}})$ and
can be derived from equation~(\ref{eq:e_r}) with $\varphi$ set to the
angle between ${\mathbf{\theta}}$ and the positive x-axis. We can then
define the rotated ellipticity correlation functions (Miralda-Escud\'e
1991, Kaiser 1992)
\begin{eqnarray}
\label{eq:e_corr}
C_{1}(\theta) & = & \langle \epsilon_{1}^{r}(0) 
  \epsilon_{1}^{r}({\mathbf{\theta}}) \rangle \nonumber \\
C_{2}(\theta) & = & \langle \epsilon_{2}^{r}(0)
  \epsilon_{2}^{r}({\mathbf{\theta}}) \rangle,
\end{eqnarray}
where the brackets denote an average over pairs of galaxies separated
by an angle $\theta$. The correlation functions $\langle
\epsilon_{1}^{r}(0) \epsilon_{2}^{r}({\mathbf{\theta}}) \rangle$ and
$\langle \epsilon_{2}^{r}(0) \epsilon_{1}^{r}({\mathbf{\theta}})
\rangle$ are expected to vanish since they flip sign under a parity
transformation ($x \rightarrow -x, y \rightarrow y$). In the following
section, we compute the amplitude of these correlation functions
expected from weak lensing. The predictions for $C_{1}(\theta)$ and
$C_{2}(\theta)$ from intrinsic correlations will be presented in
\S\ref{intrinsic}.

\section{Correlations from weak lensing}
\label{lensing}
Weak gravitational lensing produces coherent distortions in the images
of background galaxies (see Mellier 1999; Kaiser 1999; Bartelmann \&
Schneider 1999 for recent reviews). This effect is characterized by
the distortion matrix $\Psi_{ij} \equiv \partial (\delta \theta_{i}) /
\partial \theta_{j}$ where $\delta \theta_{i}(\theta_{j})$ is the
angular displacement field induced by lensing at position
$\theta_{j}$. The trace free part of the distortion matrix is called
the shear $\gamma_{i} \equiv \{ \Psi_{11} - \Psi_{22}, 2 \Psi_{12} \}
/ 2$ and can be directly measured from the ellipticity of background
galaxies (if they are intrinsically uncorrelated). For the definition
of ellipticity in Equation~(\ref{eq:e_def}) and in the weak lensing
regime, the shear is indeed related to the average galaxy ellipticity
by e.g. Rhodes et al. (1999).
\begin{equation}
\gamma_{i} = \langle \epsilon_{i} \rangle / \lambda, 
\end{equation}
where the brackets denote an average over randomly oriented galaxies,
$\lambda \equiv 2 (1-\sigma_{\epsilon}^{2})$, and $\sigma_{\epsilon}^{2}
\equiv \langle \epsilon_{1}^{2} \rangle = \langle \epsilon_{2}^{2}
\rangle$ is the ellipticity variance of the galaxies in the absence of
lensing. In the following, we will adopt $\sigma_{\epsilon} \simeq
0.3$ as is typically found in weak lensing surveys (e.g. Rhodes et al.
1999; Bacon et al. 2000), yielding $\lambda \simeq 1.8$.

For weak lensing, the ellipticity correlation functions
(Eq.~[\ref{eq:e_corr}]) are given by (Miralda-Escud\'e 1991, Kaiser
1992, Kamionkowski et al. 1997; Bacon,
Refregier \& Ellis 2000)
\begin{equation}
C_{i}(\theta) = \frac{\lambda^{2}}{4\pi} \int_{0}^{\infty} dl~l
C^{\gamma}_{l} \left[ J_{0}(l\theta) + (-1)^{i+1} J_{4}(l \theta)
\right],
\end{equation}
where $C^{\gamma}_{l}$ is the shear power spectrum given by
(e.g. Bacon, Refregier \& Ellis 2000)
\begin{equation}
C^{\gamma}_{l} = \frac{9}{16} \left( \frac{H_{0}}{c} \right)^{4}
 \Omega_{m}^{2} \int_{0}^{\infty} d\chi~
 \left[ \frac{g(\chi)}{ar(\chi)} \right]^{2} P\left(\frac{l}{r},\chi
 \right),
\end{equation}
where $P(k,\chi)$ is the 3-dimensional mass power spectrum at comoving
radius $\chi$ and $a$ is the scale factor normalised to unity
today. The comoving angular-diameter distance is $r(\chi)=R_{0}
\sinh(\chi R_{0}^{-1})$, $\chi$, and $R_{0} \sin(\chi R_{0}^{-1})$, in
an open, flat and closed universe respectively. The scale radius at
present is $R_{0}=c/(\kappa H_{0})$, with $\kappa^{2}=1-\Omega$, $1$,
and $\Omega-1$, in each case respectively. The radial weight function
$g(\chi)$ is given by
\begin{equation}
g(\chi) = 2 \int_{\chi}^{\infty} d\chi' p_{\chi}(\chi')
\frac{r(\chi) r(\chi'-\chi)}{r(\chi')}.
\end{equation}
The selection function $p_{\chi}(\chi)$ is the probability that an
object at radius $\chi$ is included in the catalogue and is normalised
as
\begin{equation}
\label{eq:p_norm}
\int d \chi~ p_{\chi}(\chi) \equiv 1.
\end{equation}
It is related to the redshift probability function $p_{z}$ by
$p_{\chi} d\chi = p_{z} dz$. We will consider a redshift distribution
of the form
\begin{equation}
\label{eq:z_dist}
p_{z}(z) \propto z^{2} \exp\left[-\left(\frac{z}{z_{0}}\right)^{\beta}\right],
\end{equation}
which gives an approximate description of the observed distribution
for $\beta \simeq 1.5$ (e.g. Smail et al. 1995). The mean and median
redshift of this distribution are $\langle z \rangle \simeq 1.5 z_{0}$
and $z_{m} \simeq 1.4 z_{0}$, respectively. We will consider two
distributions with $z_{m} = 1.0$ and $z_{m} \simeq 0.2$,
corresponding, respectively, to that for current weak lensing surveys
and for wide but shallower surveys such as SuperCOSMOS, APM and SDSS
(Maddox et al. 1990, Gunn \& Weinberg 1995).

\begin{figure}
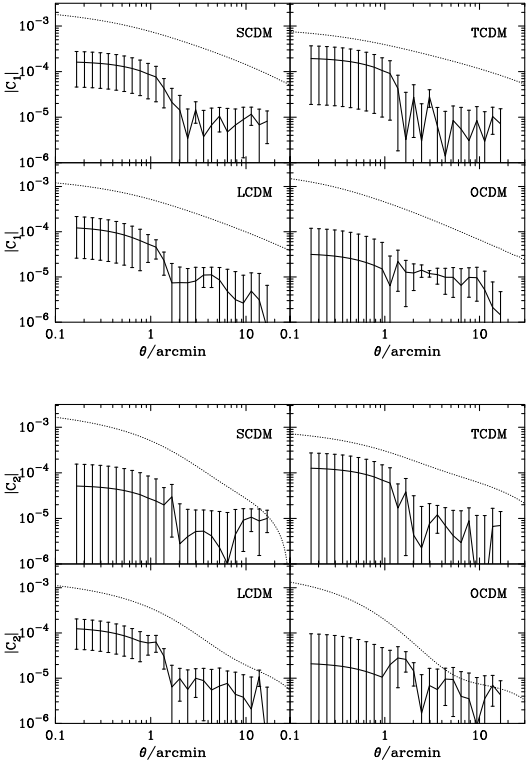

\psfig{figure=fig2a.ps,width=78mm,angle=270} 
\psfig{figure=fig2b.ps,width=78mm,angle=270} 
\caption{Intrinsic ellipticity correlation functions $|C_{1}(\theta)|$ 
(top group) and $|C_{2}(\theta)|$ (bottom group) for spiral galaxies,
for each cosmological model.  The correlation functions expected
from lensing are shown as dotted lines. The source redshift was taken
to be $z_{s}=1$, corresponding to current dedicated weak lensing
surveys.  Note that error bars are correlated.}
\label{fig:cth_li}
\end{figure}

Figure~\ref{fig:cth_li} (dotted lines) shows the correlation functions
expected for lensing for the cosmological models of
Table~\ref{tab:models}. They were derived using the fitting formula of
Peacock and Dodds (1996) for the non-linear evolution of the mass
power spectrum. The redshift distribution of the sources was taken to
be that of Equation~(\ref{eq:z_dist}) with $z_{m}=1$. Note that these
correlation functions have a specific angular dependence which can be
used as a signature of lensing. In particular, $C_{2}$ turns negative
for $\theta \ga 20'$ (not shown) for all models (Miralda-Escud\'e
1991, Kaiser 1992, Kaiser 1998; Kamionkowski et al. 1998).

\begin{table}
 \centering \begin{minipage}{140mm} \caption{Cosmological models}
 \label{tab:models} \begin{tabular}{crrrr} \hline Model & $\Omega_{m}$
 & $\Omega_{\Lambda}$ & $\sigma_{8}$ & $\Gamma$ \\ \hline SCDM & 1.0 &
 0 & 0.51 & 0.50 \\ $\tau$CDM & 1.0 & 0 & 0.51 & 0.21 \\ LCDM
 & 0.3 & 0.7 & 0.90 & 0.21 \\ OCDM & 0.3 & 0 & 0.85 & 0.21 \\ \hline
\end{tabular}
\end{minipage}
\end{table}

\begin{figure}
\psfig{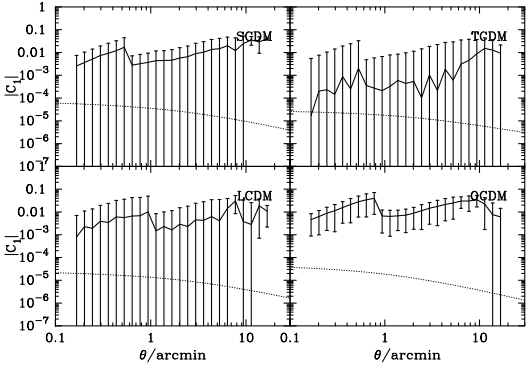} 

\caption{Same as the top group of the previous
figure, but for a median galaxy redshift of $z_{m}=0.2$, as
appropriate for wide shallower surveys such as SDSS.  
%Graph is shown only for $C_1$ for LCDM; other plots are similar.  
We have too few small-separation pairs to constrain the intrinsic
correlations between close pairs on the sky in a shallow survey. Note
that error bars are correlated; most of the estimate comes from the
first two bins in 3D, and the final panel is not a significant
detection, since it is a log scale.}
\label{fig:cth_li_near}
\end{figure}

Figure~\ref{fig:cth_li_near} shows an example of the lensing correlation
function for a median redshift of $z_{m}=0.2$.  We see
that the amplitude is more than one order of magnitude lower than
that for $z_{m}=1$. This reflects the fact that lensing produces
coherent ellipticities throughout the depth of the survey.  It is
therefore advantageous to use deep surveys to detect lensing.  This is
in contrast to intrinsic correlations which are expected be important
on small spatial scales, and thus be diluted when averaged over the
depth of deep lensing surveys.  This is reflected to a certain extent
in the figure, but we do not have enough pairs at small physical
separation to constrain the intrinsic signal for shallow surveys.

\section{Intrinsic Correlations}
\label{intrinsic}

\subsection{Galaxy Models}

We have considered two galaxy models, `ellipticals' and `spirals',
based on halos identified in N-body simulations.  Our elliptical model
is similar to that considered by Croft \& Metzler (2000) and assumes
that the ellipticity of the galaxy is the same as the ellipticity of
the dark matter halo; since the halos contain rather few particles
(10-146), we are not confident that the true halo ellipticity is
computed accurately, due to possible numerical artefacts.  We thus
concentrate on a model for spiral galaxies, for which we use the
angular momentum of the halo, not its shape.  The fraction of spirals
in the field population is high, although observationally
sample-dependent (e.g. Loveday et al. 1992, Lin et al. 1996, Folkes et
al. 1999), which also motivates our choice.

\begin{figure}
\psfig{figure=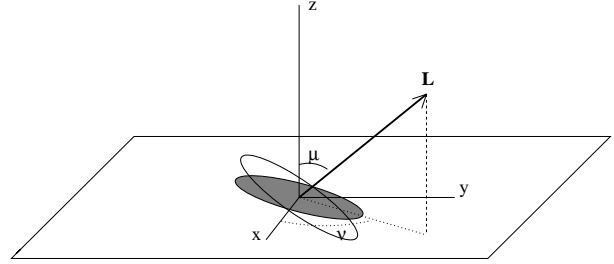,width=80mm} 

\caption{Simple model of a spiral galaxy which is taken to be 
a thin disk perpendicular to its angular momentum vector ${\mathbf
L}$.  The coordinate system is chosen so that the sky is in the
$x$-$y$ plane and the line of sight is along the $z$-axis. The disk of
the galaxy is shown as the open ellipse, and its projection on the sky
as the filled ellipse.}
\label{fig:disk}
\end{figure}

We model a spiral galaxy as a thin disk which is assumed to be
perpendicular to the angular momentum vector ${\mathbf L}$ of its
parent halo. The geometry of such a disk is shown in
Fig.~\ref{fig:disk}. Let us choose a coordinate system such that the
sky is in the $x$-$y$ plane, and the line of sight is along the
$z$-axis. Let us call the polar angle and the azimuthal angle of
${\mathbf L}$ in this coordinate system as $\mu$ and $\nu$,
respectively. The disk is shown as the open ellipse on the figure, and
its projection on the sky as the filled ellipse.  It is easy to show
that the ellipticity $\epsilon \equiv ( \epsilon_{1}^{2} +
\epsilon_{2}^{2} )^{\frac{1}{2}}$ and position angle $\alpha$ of the
projected ellipse (eq.~[\ref{eq:e_def}]) are given by
\begin{eqnarray}
\label{eq:disk}
\epsilon & = & \frac{\sin^{2} \mu}{1+\cos^{2} \mu}, \nonumber \\
\alpha & = & \nu + \frac{\pi}{2}.
\end{eqnarray}
Note that the observed ellipticity $\epsilon_{i}$ depends only on the
orientation of ${\mathbf L}$, and not on its magnitude; it is also
independent of the surface brightness profile, provided it depends
only on radius.  The `elliptical' model assigns ellipticities from the halo
quadrupole moments as discussed in Kaiser \& Squires (1993).

\begin{figure}
\psfig{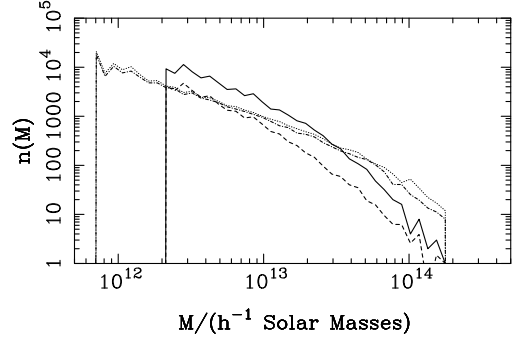} 
\caption{Mass histograms for the halos in the 4 model boxes, with 50
bins equally-spaced logarithmically.  Minimum mass
is set by requiring halos to have at least 10 particles.  The maximum
mass we consider in the analysis is $10^{13} h^{-1}M_\odot$.}
\label{fig:nm}
\end{figure}

\subsection{Simulations}

\begin{figure}
\psfig{figure=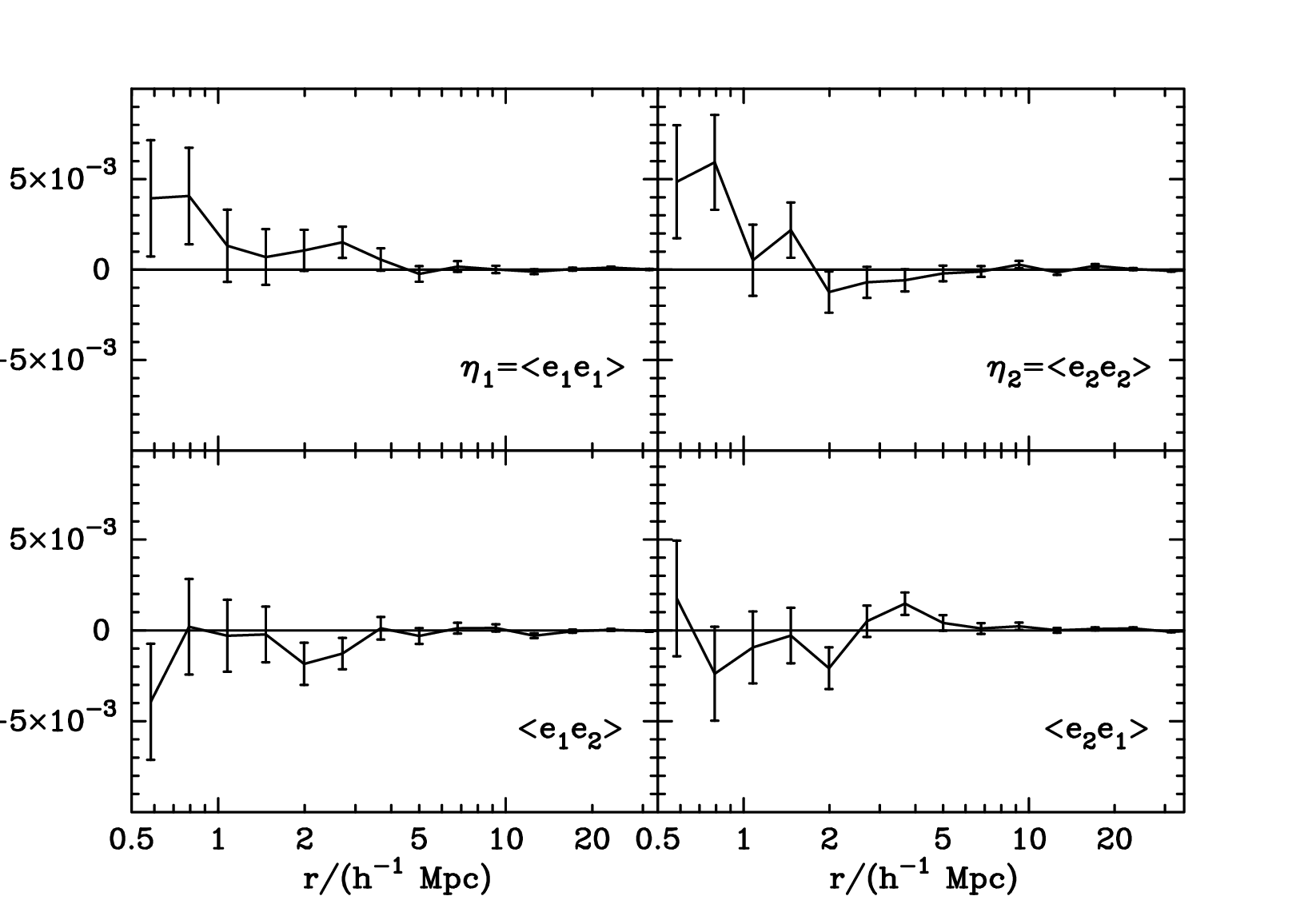,width=80mm,angle=270} 
\caption{Three-dimensional correlation functions for spirals in the
$LCDM$ model at a redshift $z=1$.} 
\label{fig:corr3ds}
\psfig{figure=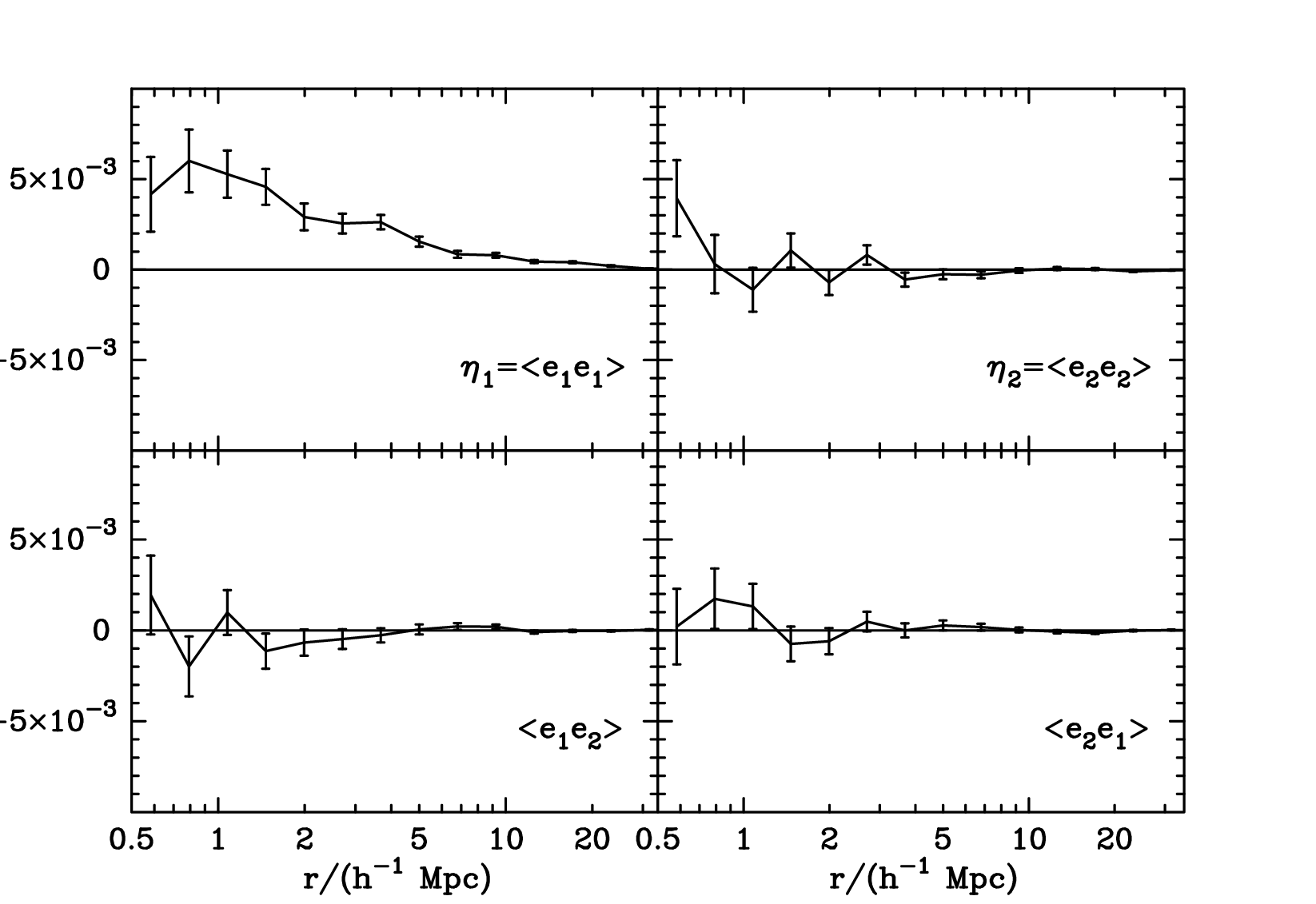,width=80mm,angle=270} 
\caption{Three-dimensional correlation functions for ellipticals in the
$LCDM$ model at a redshift $z=1$.} 
\label{fig:corr3de}
\end{figure}

We use N-body simulations developed by the Virgo Consortium (Jenkins
et al. 1998) for the 4 cosmological models with parameters listed in
table~\ref{tab:models}.  The simulations have $256^3$ particles, box
sizes of 239.5 $h^{-1}$ Mpc, and use a parallel, adaptive
particle-particle, particle-mesh scheme ($h\equiv H_0/100
km\,s^{-1}\,Mpc^{-1}$).  We have one simulation per cosmological
model.  Halos are found using a friends-of-friends algorithm, with a
linking length of 0.2 times the mean particle separation.  In each
simulation, we identify about $3 \times 10^4 - 1.2 \times 10^{5}$ dark
matter halos, dependent on the model, containing at least 10
particles, and with masses under $10^{13} h^{-1} M_\odot$ and measure
their angular momenta.  The minimum mass is $6.9 \times 10^{11} h
M_\odot$ for low-density models and $2.3
\times 10^{12} h M_\odot$ for high-density models.  Mass distributions 
are shown in Fig. \ref{fig:nm}.  We choose one side of the box as the plane of
the sky and compute the ellipticity of a disk at right-angles to the
halo angular momentum, using Equation~(\ref{eq:disk}).  Since we
assume the disks to be thin, the magnitude of the angular momentum is
irrelevant.  There is a possibility of angular momentum transfer
between the baryon component and the halo, through sub-clumps coupling
to the halo potential (e.g. Navarro \& Steinmetz 1997), but we assume
no change in the direction of {\bf L}. We then select pairs of halos
and compute the 3-dimensional ellipticity correlation functions
\begin{eqnarray}
\eta_{1}(r) & \equiv & \langle \epsilon_{1}^{r}(0) \epsilon_{1}^{r}(r)
  \rangle \nonumber \\
\eta_{2}(r) & \equiv & \langle \epsilon_{2}^{r}(0) \epsilon_{2}^{r}(r)
  \rangle,
\end{eqnarray}
where the superscript $^{r}$ denotes rotated ellipticities
(Eq.~[\ref{eq:e_r}]), and $r$ is the comoving distance between the two
galaxies.  Strictly speaking, the correlation functions depend on the
orientation of the radius vector with respect to the line-of-sight,
but we average over this angular dependence.  For a projected
catalogue, this averaging will be a good approximation for small
separations ($\ll$ the scale over which the selection function
changes), as the pairs in the catalogue will be distributed
more-or-less isotropically.  Figures \ref{fig:corr3ds} and
\ref{fig:corr3de} show the resulting 3-dimensional correlation
functions for the LCDM model at redshift 1, for spirals and
ellipticals respectively. A small but significant signal is detected
on scales smaller than a few Mpc. The other models yield similar
results.  A useful test is that $\langle e_1^r(0)e_2^r(r)\rangle$ and
$\langle e_2^r(0)e_1^r(r)\rangle$ are always consistent with zero
(lower panels), as demanded by parity considerations.  In the next
section, we use these results to compute the projected ellipticity
correlation function.

\subsection{Ellipticity Correlation Functions}

Since the halo 3D correlation functions ($\eta_{1}$ and $\eta_{2}$),
are rotationally invariant by construction, we can compute the
observed 2-dimensional intrinsic correlation functions $C^{\rm
int}_{1}$ and $C^{\rm int}_{2}$ (Eq.~[\ref{eq:e_corr}]) by
integration.  Since the observed correlation functions are
pair-weighted, we obtain
\begin{equation}
C^{\rm int}_{i}(\theta) \simeq \frac {\int_{0}^{\infty} d \chi_{1}
\int_{0}^{\infty} 
d\chi_{2} ~p_{\chi}(\chi_{1}) p_{\chi}(\chi_{2}) [ 1 + \xi(r_{12}) ]
\eta_{i}(r_{12})}{\int_{0}^{\infty} d \chi_{1} \int_{0}^{\infty} d
\chi_{2} ~p_{\chi}(\chi_{1}) p_{\chi}(\chi_{2}) [ 1 + \xi(r_{12}) ]}
\label{e_angular}
\end{equation}
where $\chi$ is the comoving radius, and $\xi(r)$ is the spatial
correlation function of the galaxy positions.  We assume $\xi(r)=(r/5
h^{-1} {\rm Mpc})^{-1.8}$ and ignore evolution, in view of the weak
evolution seen in galaxy samples since a redshift of 3 (Giavalisco et
al. 1998).  Similarly, the 3D ellipticity correlation will in general
evolve with time.  In fact we see little difference with epoch.  For
the lensing of a background with median redshift $z_m=1$, we take the
halos from simulations at $z=1$.  For the shallower samples with
$z_m=0.2$, we use the correlation function from simulations at $z=0$.

Equation (\ref{e_angular}) could be simplified to
something similar to Limber's equation for the angular correlation
function (Limber 1953), but we simply integrate this expression, using
the small angle approximation so the comoving distance $r_{12}$
between the two galaxy positions is
\begin{equation}
r^{2}_{12} \simeq (\chi_{1}-\chi_{2})^2 + 
r^{2}\left(\frac{\chi_{1}+\chi_{2}}{2}\right)\theta^{2},
\end{equation}
where $r(\chi)$ is the comoving angular-diameter distance defined in
\S\ref{lensing}.  For the $\Lambda$ model, we use $r$ derived from a 
fitting formula in Pen (1999).  The selection function $p_{\chi}(\chi)$ is 
that defined in equation~(\ref{eq:z_dist}).  Specifically, we use the
3D ellipticity correlation function as computed from the simulations,
and linearly-interpolate between points.  

\begin{figure}
\psfig{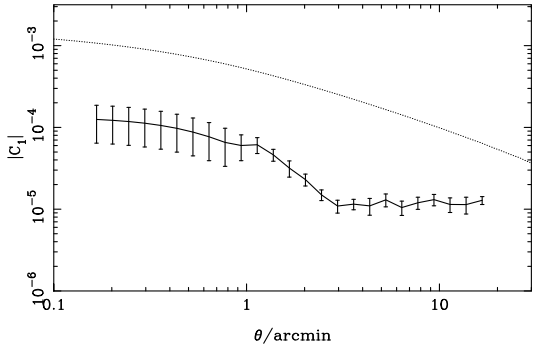} 
\caption{Intrinsic ellipticity correlation function (solid) for
ellipticals in the $LCDM$ model at a redshift $z=1$, along with the
expected signal from weak lensing (dotted).}
\label{fig:angular_ell}
\end{figure}

Figure~\ref{fig:cth_li} shows the resulting projected correlation
functions for $z_{m}=1$ for each model.  We generate 50 realisations
of the 3D correlation function, using the computed errors, and
project, using equation (\ref{e_angular}).  The resulting errors in
the angular correlation function will be correlated.  The case of
$z_{m}=0.2$ is shown in Fig. ~\ref{fig:cth_li_near}.  Croft \& Metzler
(2000), using essentially the `elliptical' model, present very similar
results, using the LCDM model and higher-resolution, smaller
simulations in addition to that used here.  In
Fig. \ref{fig:angular_ell}, we present elliptical results only for the
LCDM model with $z_m=1$, which is the model they consider.  Note that
the elliptical correlations are less noisy than the spirals.  They use
higher-resolution simulations and a different algorithm for the
halo-finder, and a slightly different selection function, but their
results are very similar to ours.  It is interesting to note that the
spiral and elliptical models give very similar correlation functions.
This is surprising, as there is a large scatter between the elliptical
and spiral ellipticity parameters.

\section{Impact for weak lensing surveys and detectability}
\label{impact}
The impact of intrinsic correlations on current weak lensing surveys
(with $z_{m}\sim1$) can be established by examining
figure~\ref{fig:cth_li}. They reveal that the intrinsic correlation
functions are generally below the lensing correlation functions for
$\theta \la 10'$. However, they are not far below for some models, and
our results indicate that intrinsic shape correlations could be
non-negligible.

It is interesting to note that Van Waerbeke et al. (2000) found
evidence for an ellipticity variance signal in excess of lensing on
$0.5' \la \theta \la 3.5'$ scales in their cosmic shear survey. This
excess signal vanishes if galaxy pairs with separations less than
$10''$ are removed. They interpret this signal as arising from
overlapping isophotes, but do not dismiss the possibility that it is
due to intrinsic correlations. Our results indicate that the latter
explanation, although possible, is unlikely.  (Note that they use a
slightly different statistic, but their level of excess is in excess
of our $\sqrt{C_{1,2}}
\sim 0.01$).

Wider but shallower surveys such as SuperCOSMOS, APM and SDSS (Maddox
et al. 1990; Gunn \& Weinberg 1995) can also be used to search for
weak lensing (Stebbins, McKay \& Frieman 1995).
Figure~\ref{fig:cth_li_near} shows the impact of intrinsic
correlations on such a survey (with $z_{m} \sim 0.2$).  In this case,
the limits on the intrinsic correlations are much larger than the
expected lensing signal. This can be understood in simple terms: while
lensing produces correlations which are coherent through the depth of
the survey, intrinsic shapes are correlated in a localised region of
space. As a result, while the amplitude of lensing is smaller for
shallower surveys, that for intrinsic correlations is larger, since it
is not diluted as in the case of deeper surveys. Intrinsic
correlations could thus be important for shallow surveys. In fact,
depending on its level, and the degree of evolution, one might be able
to use shallow surveys to measure the the intrinsic correlation
effect, and then use these results to validate or correct the
correlation function derived from deep lensing surveys.  Of course,
the ellipticity correlation may be of intrinsic interest, possibly
acting as a discriminator between cosmological models, although this
looks ambitious at present.

\section{Conclusions}
\label{conclusion}
We have studied the effect of intrinsic shape correlations on weak
lensing surveys. This effect can be thought as being analogous to the
foreground contributions to the anisotropy of the Cosmic Microwave
Background: it must be accounted for in the interpretation of weak
lensing surveys, and can provide important cosmological information
in itself.

We have considered two models for the galaxies.  Most of our results
are based on a model for spiral galaxies, where the galaxy is assumed
to be a thin disk perpendicular to the halo angular momentum.  We also
show some results for an `elliptical' model, where we assumed that the
ellipticity of the galaxy is the same as the ellipticity of the halo.
The relatively small number of particles in the halos leads to some
uncertainty in the shape parameters of the halo, so we have therefore
concentrated on the spiral model.  We measured the resulting projected
ellipticity correlation function arising from spin-spin correlation of
$30,000 - 10^{5}$ halos in an N-body simulations, for several CDM
models.  The correlations we find lie between $\sim 10^{-4}$ at small
separations to $\sim 10^{-5}$ on scales of $10'$, and may be
explicable in terms of the statistics of the initial gravitational
potential (Catelan, Kamionkowski and Blandford 2000).  The intrinsic
correlations are generally below the expected lensing signal for deep
surveys.  We note that other non-gravitational effects could increase
the correlations, through tidal interactions forming tidal tails, for
example.  However, since these effects would be confined to small
separations in 3D, our feeling is that they are unlikely to be of
importance in the angular correlations, especially for the deep
samples.

For a survey with galaxies at a median redshift $z_{m}=1$, such as
current weak lensing surveys, intrinsic correlations lie below the
lensing signal on angular scales $\theta \la 10'$, a scale where the
lensing signal is $\sim 10^{-5}$. The effect appears to be too small
to explain the excess power found on small scales by Van Waerbeke et
al. (2000).

While lensing produces correlations which are coherent over the depth
of the survey, intrinsic shape correlations are only important over
limited physical separations. As a result, intrinsic correlations are
diluted when integrated over the depth of a deep, dedicated lensing
survey. On the other hand, wider but shallower surveys such as
SDSS ($z_{m}\sim 0.2$) will be much more sensitive to intrinsic
correlations. We cannot exclude the possibility that intrinsic
correlations could be comparable to or even dominate the weak lensing
signal on all scales for such surveys.  Caution must thus be exerted
when interpreting the weak lensing signal from these surveys. With
sufficient signal, the lensing signal can be secured using the
specific angular dependence of its induced ellipticity correlations,
but shot noise will be a strong limiting factor at scales of an
arcminute or less in a survey like SDSS (see Munshi and Coles 2000 for
further discussion of errors). 

Intrinsic shape correlations are interesting in their own right, as
they provide a probe to the generation of angular momentum during
galaxy formation. Galaxy spins can for instance be used, in principle,
to measure the shear of the density field (Lee \& Pen 2000). The
SuperCOSMOS, APM and SDSS surveys can thus be used to measure the
intrinsic correlation functions with high accuracy. Intrinsic
correlations can also be constrained using 2-dimensional galaxy-galaxy
lensing to measure the alignment of mass and light in galaxies
(Natarajan \& Refregier 2000). These techniques can then be used to
constrain models of galaxy formation, and to secure the interpretation
of deeper weak lensing surveys.

\section*{Acknowledgments}

The simulations analysed in this paper were carried out using data
made available by the Virgo Supercomputing Consortium
(http://star-www.dur.ac.uk/~frazerp/virgo/) using computers based at
the Computing Centre of the Max-Planck Society in Garching and at the
Edinburgh Parallel Computing Centre.  We are very grateful to Rob
Smith for providing halos from the simulations.  We thank Rachel
Somerville, Priya Natarajan and Rob Crittenden for useful discussions,
and the referee for helpful comments. AR was supported by a TMR
postdoctoral fellowship from the EEC Lensing Network, and by a Wolfson
College Research Fellowship. Some computations used Starlink
facilities.

\bsp

\label{lastpage}


\begin{thebibliography}{99}
\bibitem{bac00} Bacon D., Refregier A., Ellis R., 2000,
 submitted to MNRAS, preprint astro-ph/0003008
\bibitem{bar99} Bartelmann M., Schneider P., 1999, submitted to
  Physics Reports, preprint astro-ph/9912508
\bibitem{Ber97}Bernardeau F., Van Waerbeke L., Mellier Y., 1997, A\& A, 
322, 1
\bibitem{Bland91} Blandford R.D., Saust A.B., Brainerd T.G., Villumsen J.V., 
1991, MNRAS, 251, 600 
\bibitem{CT96} Catelan P.,  Theuns T., 1996a,  MNRAS, 282, 436
\bibitem{CKB00} Catelan P.,  Kamionkowski M., Blandford R.D., 2000, astroph 
0005470
\bibitem{CNPT00} Crittenden R., Natarajan P., Pen U., Theuns T.,
  2000, in preparation
\bibitem{CM00} Croft R.A.C., Metzler C.A., 2000, astroph/0005384
\bibitem{D70} Doroshkevich A.G., 1970,  Afz, 6, 581
\bibitem{Folkes}{Folkes S., et al, 1999, MNRAS, 308, 459}
\bibitem{b6} Fort B., Mellier Y., 1994, Astron. Astr. Rev. 5, 239.
\bibitem{Gia98} Giavalisco M., Steidel C.C., Adelberger K.L.,
 Dickinson M.E., Pettini M., Kellogg M., 1998, ApJ, 503, 543
\bibitem{HP88} Heavens A.F., Peacock J.A., 1988,  MNRAS, 232, 339
\bibitem{hoe99} Hoekstra H., Franx M., Kuijken K., 1999,
  in {\it Gravitational Lensing: Recent Progress and Future Goals},
  ASP conference series, eds. Brainerd T., Kochanek C.S.
\bibitem{Hoyle} Hoyle F., 1949, in Burgers J. M., van de Hulst H. C.,
  eds., in  Problems of Cosmical Aerodynamics, Central Air Documents,
   Dayton, Ohio, p. 195
\bibitem{GW95}Gunn J.E., Weinberg D.H., 1995, in Maddox S., 
Aragon-Salamanca A., eds., in Wide Field Spectroscopy and
the Distant Universe, the $35^{th}$ Herstmonceux Conference", p.3
\bibitem{b8} Hu W., Tegmark M., 1998 (astro-ph/9811168)
\bibitem{b9} Jain B., Seljak U., 1997 ApJ, 484, 560
\bibitem{jen98} Jenkins A., et al., 1998, ApJ 499, 20
\bibitem{b12} Kaiser N., 1992, ApJ 498, 26.
\bibitem{KS} Kaiser N., Squires G., 1993, ApJ 404, 441.
\bibitem{b13} Kaiser N., 1998, ApJ 498, 26.
\bibitem{kai98} Kaiser N. et al., 1998, submitted to ApJ, preprint
  astro-ph/9809268
\bibitem{kai99a} Kaiser N., 1999a, Review talk for Boston 99 lensing
  meeting, preprint astro-ph/9912569
\bibitem{kai00} Kaiser N., Wilson G., Luppino G.A., 2000,
  submitted to ApJL, preprint astro-ph/0003338
\bibitem{kam97} Kamionkowski M., Babul A., Cress, C.M., Refregier
  A., 1997, MNRAS, 301, 1064, preprint astro-ph/9712030
\bibitem{LeePen}Lee J., Pen U.-L., 2000, ApJ, 532, L5
\bibitem{limber} Limber D.N., 1953 ApJ 117, 134.
\bibitem{APM}Maddox S.J., Sutherland W.J., Efstathiou G., Loveday J.,
1990, MNRAS, 243, 692
\bibitem{Lin}{Lin H., Kirshner R.P., Shectman S.A., Landy S.D., 
Oemler A., Tucker D.L., Schechter P.L., 1996, ApJ, 464, 60}
\bibitem{Love}{Loveday J., Peterson, B.A.,
 Efstathiou G., Maddox S.J., 1992, ApJ, 390, 338} 
\bibitem{mel99} Mellier Y., 1999, ARAA, 37, 127
\bibitem{Mir91}Miralda-Escud\'e J., 1991, ApJ, 380, 1
\bibitem{MC}Munshi D., Coles P., 2000, astro-ph/0003481
\bibitem{nat}Natarajan P., Refregier A., 2000, to appear
  in ApJL, preprint astro-ph/0003344
\bibitem{NS}Navarro J.M., Steinmetz M., 1997, ApJ, 478, 13
\bibitem{pea99} Peacock J.A., 1999, {\it Cosmological Physics},
  (Cambridge University Press: Cambridge)
\bibitem{pea96} Peacock J., Dodds S.J., 1997, MNRAS, 280, L19
\bibitem{Pee69} Peebles P.J.E., 1969,  ApJ,155, 393
\bibitem{Pen}Pen U.-L., 1999, ApJS, 120, 49
\bibitem{pen2}Pen U.-L., Lee J., Seljak U., 2000, submitted
  to ApJL, astro-ph/0006118
\bibitem{rho99} Rhodes, J., Refregier A., Groth E., 1999,
  to appear in ApJ, astro-ph/9905090
\bibitem{b21} Schneider P., 1996 in {\it Universe at High Z},
  eds. Martinez-Gonzalez E., Sanz J., p470.
\bibitem{Sch98} Schneider P., Van Waerbeke L., Jain B., Kruse G., 1998, MNRAS, 296, 893
\bibitem{sma95} Smail I., Hogg D.W., Cohen J.G., 1995, ApJ, 449, L105
\bibitem{ste95} Stebbins A., McKay T., Frieman J.A.,
  1995, in Proc. IAU Symposium 173, preprint astro-ph/9510012
\bibitem{sug99} Sugerman B., Summers F.J., Kamionkowski M., 1999,
  submitted to MNRAS, astro-ph/9909266
\bibitem{Tor}Tormen G., 1997, MNRAS, 290, 411
\bibitem{van98} Van Waerbeke L., Bernardeau F., Mellier Y.,
  1999, A\&A in press, astro-ph/9807007
\bibitem{van00} Van Waerbeke L., et al. 2000, submitted
  to A\&A, preprint astro-ph/0002500
\bibitem{Vill96} Villumsen J.V., 1996, MNRAS, 281, 369
\bibitem{WVD} West M.J., Villumsen J.V., Dekel A., 1991, ApJ, 369, 287
\bibitem{} White S.D.M., 1984,  ApJ, 286, 38
\bibitem{wit00} Wittman D.M., Tyson J.A., Kirkman D.,
  Dell'Antonio I., Bernstein G., 2000, submitted to Nature,
  preprint astro-ph/0003014
\end{thebibliography}
\end{document}